\documentclass{kapproc} 

\usepackage{amsmath}
\usepackage{amsfonts}
\usepackage{amssymb}

\usepackage[dvips]{graphicx}

\normallatexbib 

\newcommand{\Tbf}{{\bf T}}
\newcommand{\phibf}{\mathbf{\Phi}}

\begin{document}

\articletitle{Heating of oil well by hot water circulation}


\author{Mladen Jurak}
\affil{Department of Mathematics\\
University of Zagreb\\
Zagreb, Croatia}
\email{jurak@math.hr}

\author{\v Zarko Prni\'c}
\affil{INA-Naftaplin, d. o. o.\\
Zagreb, Croatia }
\email{zarko.prnic@ina.hr}

\begin{abstract}
When highly viscous oil is produced at low temperatures, large
pressure drops will significantly decrease production rate. One of
 possible solutions to this problem is  heating of oil well by
hot water recycling.
We construct and analyze a mathematical model of oil-well heating
composed  of three linear parabolic PDE coupled
with one Volterra integral equation. 
Further on we construct numerical method for the
model and present some simulation results.
\end{abstract}


\section*{Introduction}
An oil well producing at low temperatures may experience large pressure
drops due to high viscosity of oil and wax forming. One way
to avoid these pressure drops is heating of oil by hot water  recycling.

The tubing is surrounded by two annulus for water circulation. Hot
water is injected into inner annulus and  it flows out of the
system through the outer annulus. The main technical concern is
minimization of energy lost in the system while keeping oil 
temperature sufficiently high.

Configuration just described will be called counter flow exchange.
If the hot water is injected into outer annulus and leaves the system
through inner annulus, then we talk about 
 parallel heat flow exchange.

\begin{center}
\includegraphics[scale=.5]{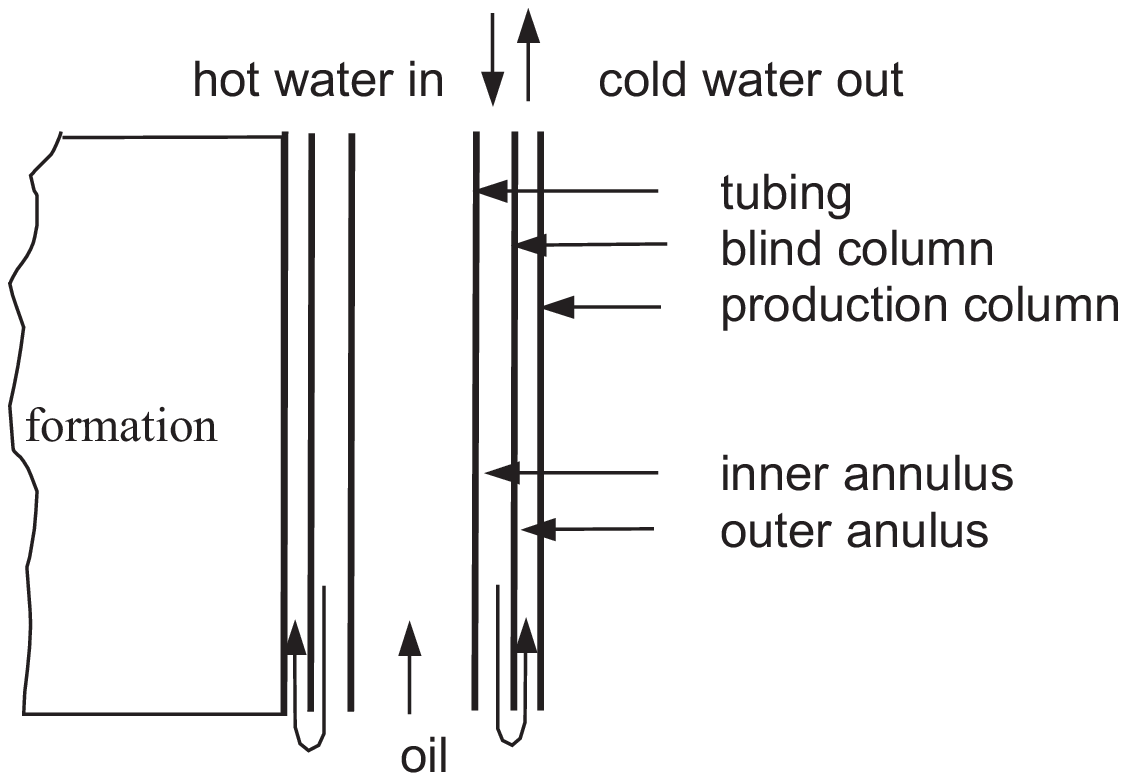}
\end{center}
\vspace{-20pt}
\begin{center}
{\small Figure 1. Counter flow heat exchange}
\end{center}
\vspace{10pt}

The outline of the paper is as follows.
In the first section we present a simple one-dimensional mathematical 
model describing the heat exchange in the system.
We present only  counter flow configuration since  parallel  flow 
configuration differs only in signs of water velocities.
Solvability of a system of integro-differential equations
describing the heat exchange 
is discussed in the second section. It is shown that the result of 
Artola \cite{artola} can be applied. 
In final section we  discuss numerical method for solution approximation
and present some numerical results for counter flow and parallel  flow 
configurations.

A problem similar to this one was considered in engineering 
literature in \cite{Rd}.

\section{Mathematical model}

Cross-sectional mean velocities of  oil and water in inner and outer 
annulus will be denoted by $v_o$, $v_i$ and $v_e$. 
They are assumed to be constant, and therefore the fluids
have constant pressure drops.
Furthermore, to simplify the model, we neglect friction and
we take mass densities $\rho_o$ (oil), $\rho_w$ (water)      
to be constant.
The heat is transferred between the tubing, inner and outer annulus and the
formation according to Newton's law.

With these simplifying assumptions and taking direction of $z$
axis vertically downwards, we obtain the following  three
parabolic equations (see \cite{CJ} for example):

\begin{align}
a_o[\frac{\partial T_o}{\partial t}-v_o \frac{\partial T_o}{\partial z}]
+ b_o(T_o-T_i) & =
D_o \frac{\partial^2 T_o}{\partial z^2} \label{eq:1}\\
a_i[\frac{\partial T_i}{\partial t}+v_i \frac{\partial T_i}{\partial z}] 
+b_o (T_i-T_o) +b_e (T_i - T_e)
&= D_i \frac{\partial^2 T_i}{\partial z^2} \label{eq:2}\\
a_e[\frac{\partial T_e}{\partial t}-v_e \frac{\partial T_e}{\partial z}] 
+b_e (T_e-T_i) +b_f (T_e-T_f) 
&= D_e \frac{\partial^2 T_e}{\partial z^2}
,\label{eq:3}
\end{align}
for $z\in (0,L)$ and $t\in (0,t_{\rm max})$. 
The main variables are the temperatures of oil, water in inner annulus,
water in outer annulus and the temperature of the formation, denoted
respectively by $T_o$, $T_i$, $T_e$ and $T_f$.
All coefficients are  constant and they have the following meaning:
$a_o=A_o\rho_o c_o$, $a_i=A_i\rho_w c_w$, $a_e=A_e\rho_w c_w$
where $A_o$, $A_i$ and $A_e$ are cross-sectional areas and $c_o$, $c_w$
are heat capacities. By $b_o$,  $b_e$ and  $b_f$  are denoted heat 
transfer coefficients from Newton's law, and by $D_o$, $D_i$ 
and $D_e$ thermal conductivities, multiplied by cross-section areas. 
In counter flow exchange all three fluid velocities are  positive. 
From mass conservation it follows $a_i v_i = a_e v_e$.

Heat flow in surrounding formation is assumed to be radial with
respect to  the tubing and to have constant (geothermal) gradient in
vertical direction.
We  denote by $T_z(z)$ geothermal temperature and by
$T_s(r,z,t)$ the temperature in the soil. Formation temperature 
$T_f$ is then given by $T_f(z,t)=T_s(r_f,z,t)$, where $r_f$ is
formation radius. The temperature  $T_s$ is the solution of the
heat equation with initial temperature $T_z$, temperature
at infinity equal to $T_z$, and prescribed heat flux $q_f$ at
$r=r_f$. In the other hand, $q_f$ is given by Newton's law
\begin{equation}
q_f = b_f (T_e-T_f). \label{formation}
\end{equation}
Then, by applying Duhamel's principle we can  represent 
formation temperature by the formula
\begin{align}
T_f(z,t) =T_z(z) + \int_0^t p(t-\tau) \frac{d}{d\tau} q_f(z,\tau)\, d\tau, 
\label{int-eq}
\end{align}
where $p(t)=P(r_f,t-\tau)/2\pi k_f$ ($k_f$ is thermal conductivity of
the soil) and $P(r,t)$ is the solution of the problem 
\begin{align}
\left\{
\begin{array}{l} 
\displaystyle { 
\frac{\rho_f c_f}{k_f} \frac {\partial P}{\partial t}= 
              \frac {1}{r} \frac {\partial}{\partial r} 
	      \Big(r \frac {\partial P}{\partial r}\Big)},\quad r>r_f,\; t>0\\[3mm]
	      P(r,0)=0,\quad r>r_f\\[1mm]
	      P(\infty,t)=0,  \quad t>0 \\[1mm] 
	      \displaystyle  -2\pi k_f {\frac {\partial P}{\partial r} }
{\Big|}_{r=r_f}=1.\label{radial}
\end{array}\right.
\end{align}
($\rho_f$ and $c_f$ are mass density and heat capacity of the soil,
respectively). It can be shown as in 
van Everdingen and Hurst \cite{EH} that $p(t)=O(\sqrt{t})$ and 
$p'(t)=O(1/\sqrt{t})$, as $t\to 0$. Therefore, $p'(t)$ is in $L^1_{\rm loc}([0,\infty))$
and we can make partial integration in (\ref{int-eq}).
Taking natural assumption  that $q_f=0$  at $t=0$
(that is $T_e=T_z$ at $t=0$) and using (\ref{formation}) 
we obtain a Volterra integral equation for $q_f$:
\begin{align}
b_f (T_e(z,t) -T_z(z)) = q_f(z,t) +
b_f \int_0^t p'(t-\tau) q_f(z,\tau)\, d\tau. \label{vol-eq}
\end{align}
This equation has the resolvent $r\in L^1_{\rm loc}([0,\infty))$ and it can be solved 
by formula (see Gripenberg, Londen and Staffanson \cite{volt}) 
\begin{equation}
q_f = b_f \big[T_e -T_z - r\star (T_e-T_z) \big],\label{qf-expl}
\end{equation}
where we have introduced  convolution operator
\[  (r*\phi)(t)= \int_0^t r(t-\tau) \phi(\tau)\, d\tau .\]
By use of (\ref{qf-expl})  and  (\ref{formation})
we can eliminate formation temperature from (\ref{eq:3}) which
is then transformed to
\begin{align}
a_e[\frac{\partial T_e}{\partial t}-v_e \frac{\partial T_e}{\partial z}] 
+b_e (T_e-T_i) +b_f (T_e-r \star T_e) 
&= D_e \frac{\partial^2 T_e}{\partial z^2} +F,\label{eq:4}
\end{align}
where $F=b_f (T_z-r \star T_z)$ is a smooth known function. 
We see that equations (\ref{eq:1}), (\ref{eq:2}) and
(\ref{eq:4}) represent parabolic system perturbed by the operator $M$ given by
\begin{equation}
M u(z,t) = \int_0^t r(t-\tau) u(z,\tau)\, d\tau  \label{op-M}
\end{equation}

The problem is to solve the system 
(\ref{eq:1}), (\ref{eq:2}) and (\ref{eq:4})  with suitable boundary
and initial conditions. 
We assume given the temperatures of entering water at $z=0$ and 
oil at $z=L$. 
At the bottom of inner and outer annulus we have equality of water
temperatures  and continuity of total thermal flux. 
Therefore, we take 
\begin{gather}
\frac{\partial T_o}{\partial z}(0,t)=0,\;\; 
\frac{\partial T_e}{\partial z}(0,t)=0 ,\;\; T_o(L,t)=T_o^L,\;\;
T_i(0,t)=T_i^0,\label{bc-1}\\
T_i(L,t)=T_e(L,t),\quad 
       D_i \frac{\partial T_i}{\partial z}(L,t)+
       D_e \frac{\partial T_e}{\partial z}(L,t)=0,
\end{gather}
for all $t>0$, where $T_o^L=T_z(L)$ and $T_i^0$ are given.  
The initial conditions are
\begin{gather}
T_o(z,0)=T_e(z,0)=T_z(z),\;\; T_i(z,0)=T_z^1(z),\label{ic-1}
\end{gather}
where function $T_z^1$ satisfies compatibility conditions
$T_z^1(0)=T_i^0$, $T_z^1(L)=T_z(L)$ and it is close to geothermal
temperature $T_z$.
All functions involved are supposed to be smooth.

\section{Variational problem}

We consider variational formulation of the problem 
(\ref{eq:1}), (\ref{eq:2}) and (\ref{eq:4}) with boundary and initial conditions 
(\ref{bc-1})--(\ref{ic-1}).
Without lose of generality we can consider homogeneous boundary conditions
$T_o^L=T_i^0=0$.

We introduce Hilbert space
\[  V=\{ (\phi_o,\phi_i,\phi_e)\in H^1(0,L)^3 \colon
 \phi_0(L)=0,\;\; \phi_i(0)=0,\;\; \phi_i(L)=\phi_e(L) \}
 \]
with the norm $\|\cdot\|$ inherited from $H^1(0,L)^3$
and bilinear forms ${\cal A}$, ${\cal B}$ and ${\cal C}$ over 
$V\times V$ defined as follows: for
$\Tbf=(T_o,T_i,T_e)$, $\phibf= (\phi_o,\phi_i,\phi_e)$ we set
\begin{gather*}
{\cal A}(\Tbf,\phibf)= {\cal A}_o(T_o,\phi_o)+{\cal A}_i(T_i,\phi_i)
+{\cal A}_e(T_o,\phi_e)+
{\cal B}(\Tbf,\phibf)\\
{\cal A}_o(T_o,\phi_o)= \int_0^L (D_o \frac{\partial T_o}{\partial z}
\frac{\partial \phi_o}{\partial z}
-a_o v_o  \frac{\partial T_o}{\partial z} \phi_o )\, dz\\
{\cal A}_i(T_i,\phi_i)= \int_0^L (D_i \frac{\partial T_i}{\partial z}
\frac{\partial \phi_i}{\partial z}
+a_i v_i  \frac{\partial T_i}{\partial z} \phi_i )\, dz\\
{\cal A}_e(T_e,\phi_e)= \int_0^L (D_e \frac{\partial T_e}{\partial z}
\frac{\partial \phi_e}{\partial z}
-a_e v_e  \frac{\partial T_e}{\partial z} \phi_e )\, dz,
\end{gather*}
\begin{align*}
{\cal B}(\Tbf,\phibf)& =
b_o\int_0^L(T_o-T_i)(\phi_o-\phi_i)\, dz +
b_e\int_0^L(T_i-T_e)(\phi_i-\phi_e)\, dz\\
&+ b_f\int_0^L T_e\phi_e\, dz \nonumber\\
{\cal C}(\Tbf,\phibf)& =- b_f\int_0^L  (r\star T_e) \phi_e\, dz.
\nonumber
\end{align*}
Duality between $V'$ and $V$ will be given by the formula
\[ \langle {\bf F},\phibf\rangle =
a_o \langle {F}_o,\phi_o \rangle +
a_i \langle {F}_i,\phi_i \rangle +
a_e \langle {F}_e,\phi_e \rangle
\]
where ${\bf F}\in V'$ is of the form ${\bf F}=(F_o,F_i,F_e)$,
$F_o,F_i,F_e\in (H^1(0,L))'$, and brackets at the right hand side
signify duality between $(H^1(0,L))'$ and $H^1(0,L)$.
We set $H=L^2(0,L)^3$, with usual norm denoted by $|\cdot |$,
and by identifying $H$ with its dual we have
$V\subset H \subset V'$, with dense and continuous injections.
Furthermore, by $W(V,V')$ we denote the space of all functions
from $L^2(0,t_{\rm max};V)$ with time derivative in $L^2(0,t_{\rm max};V')$. It is well
known that $W(V,V')$ is continuously embedded in $C([0,t_{\rm max}];H)$.

With this notations we can reformulate the problem  
(\ref{eq:1}), (\ref{eq:2}), (\ref{eq:4}), (\ref{bc-1})--(\ref{ic-1}) in 
the following variational problem:
find  $\Tbf\in W(V,V')$ such that $\Tbf(0)=\Tbf^0\in H$ and
for a.e. $t\in (0,t_{\rm max})$ 
\begin{equation}
 \langle \Tbf',\phibf\rangle +{\cal A}(\Tbf,\phibf)+{\cal C}(\Tbf,\phibf)=
 \langle {\bf F},\phibf\rangle, \quad \forall \phibf\in V.\label{var-pr}
\end{equation}
The linear form on the right hand side is given by 
\begin{gather*}
\langle {\bf F},\phibf\rangle =\int_0^L F \phi_e\, dz.
\end{gather*}
and it is obviously continuous.

It is easy to see that ${\cal A}(\cdot,\cdot)$ is continuous bilinear form on
$V$ which satisfy 
\[  {\cal A}(\Tbf,\Tbf) +  \gamma |\Tbf|^2 \geq \alpha \|\Tbf\|^2,\quad \forall \Tbf\in V,
\]
with some constants $\alpha, \gamma >0$. 
Bilinear form ${\cal C}(\cdot,\cdot)$ comes from perturbation operator $M$.
It is not difficult to see that for any function 
$u\colon (0,t_{\rm max}) \to L^2(0,L)$
it holds 
\begin{equation}
\| Mu(t)\|_{L^2(0,L)} \leq \sqrt{{\cal K}(t)} 
(\int_0^t |r(t-\tau)| \|u(\tau)\|_{L^2(0,L)} d\tau)^{1/2},
\end{equation}
where ${\cal K}(t) = \int_0^t | r(\tau)|\, d\tau $. From here it follows that 
$M$ is linear and continuous operator from $L^\infty(0,t_{\rm max};L^2(0,L))$
to itself, and it has  the following continuity property: if $u_n, u\in 
L^\infty(0,t_{\rm max};L^2(0,L))$ are such that 
\[ u_n(t)\to u(t)
   \quad\text{in } L^2(0,L)\;\;\text{for a.e. } t\in (0,t_{\rm max})
   \]
   then 
\[ Mu_n(t)\to Mu(t)
   \quad\text{in } L^2(0,L)\;\;\text{for a.e. } t\in (0,t_{\rm max}).
   \]
Furthermore, it is easy to see that $M$ is an operator of {\sl local type},
as defined in Artola \cite{artola}, and therefore we can apply 
Theorem 1 from  \cite{artola} and conclude:
\begin{theorem}
Variational problem (\ref{var-pr}) 
has a unique solution $\Tbf\in W(V,V')$ for any $\Tbf^0\in H$
and  ${\bf F}\in L^2(0,t_{\rm max};V')$.
\end{theorem}

\section{Numerical approximation}

In this section we discuss numerical approximation by finite
difference method of the problem 
(\ref{eq:1}), (\ref{eq:2}), (\ref{eq:4}), (\ref{bc-1})--(\ref{ic-1}).
Instead of using equation (\ref{eq:4}) we find more convenient to  apply finite 
difference method to the equations
(\ref{eq:1}), (\ref{eq:2}), (\ref{eq:3})  and to discretize directly
equations (\ref{formation}) and (\ref{int-eq}). We avoid numerical 
resolution of problem (\ref{radial}) by the use of  Hasan and Kabir \cite{HK}
approximation:
\[ p(t)=p_n(\frac{k_f t}{\rho_f c_f r_f^2})\]
where
\[
    p_n(s)=\begin{cases}
            \frac{2}{\sqrt{\pi}} \sqrt{s} ( 1-0.3 \sqrt{s}) & \text{for } s\leq 1.5\\
	    \frac{1}{2}(0.80907+\log(s))\left( 1+\frac{0.6}{s}\right) &
	    \text{for } s > 1.5.
         \end{cases}
\]

Furthermore, in our problem 
constants $D_o$, $D_i$ and $D_e$ are very small and it is natural
to consider hyperbolic system ($D_o=D_i=D_e=0$) instead of parabolic one.
Due to limited space we will not enter here into discussion of existence theory 
for hyperbolic system.  
We  just note that any difference scheme adapted to hyperbolic version 
of the system (\ref{eq:1})--(\ref{eq:3}) will produce certain amount of 
numerical dispersion  that will {\em cover} thermal diffusion in 
equations (\ref{eq:1})--(\ref{eq:3}), at least for reasonable mesh sizes.
Therefore we chose to neglect thermal diffusion and consequently to drop 
superfluous Neumann boundary conditions for oil and water in outer annulus. 
This will generally change the solution just in corresponding boundary
layers. 

We apply explicit finite difference scheme of first order with 
convective terms treated by {\em upwinding}. 
In all the experiments we have used a uniform grid in space and
time. The spatial step $h$ and time step $\tau$ are related by the
fixed positive number $\lambda$   through relation
$\lambda=\tau/h$.   

In discretization of integral equation (\ref{int-eq})
we use composite trapezoidal rule which gives  the following
procedure for calculation of formation temperature at $t=n\tau$ and
$z=i h$:
\begin{align*}
T_{F,i}^n = \frac{1}{1+ P_1}
\left( T_{Z,i} + 
\sum_{k=1}^{n-1} ( T_{V,i}^k - T_{F,i}^k )(P_{n+1-k}
-P_{n-1-k}) + P_{1}T_{V,i}^n \right). 
\end{align*}
As a consequence of the convolution in formula (\ref{int-eq})
we see that the solution on next time level includes the solutions on 
all previous time levels. 

It can be shown that that described explicit scheme is 
TVB  (total variation bounded) and $L^{\infty}$-stable if the following 
CFL condition is satisfied:
\[
\lambda \le \frac{1}{\max\{v_o, v_i, v_e\}+ C h},
\]
where $C>0$ is certain constant that can be calculated from the 
coefficients in (\ref{eq:1})--(\ref{eq:3}).

We now proceed with some numerical results.
To evaluate the merits of one flow arrangement over another
(counter flow and parallel flow), some
conditions must be equal. The interval of time during which the
water is cooled is not equal to the interval of time during which
the water is heated. The sum of these time intervals we call
circulating period or cycle. Both method can now be compared using
the same circulating period.

Results of our simulations after four cycles are presented in the
figure Fig. 2. Counter flow heat exchange
temperature calculations are shown on the left figure.
The tubing
temperature is almost always less then inner and greater than
outer annulus temperature.

Parallel flow  heat exchange temperature calculations are shown on the
right figure. 
The tubing temperature lies between the inner
annulus temperature and formation temperature.
In this case oil temperature is lower than any annulus temperature.
Besides, formation temperature is higher than
in the previous case.

Tubing temperature as well as outer annulus temperature reach very soon
almost constant level. The important thing to note with
respect to the bottom-hole fluid temperature is that this
temperature continually changes with time. A steady-state condition
is never attained. Hence the stabilization of both outlet
temperatures does not mean that all of the temperatures in the
circulating system are constant.

Under the same conditions we found that in  parallel-flow arrangement
temperature drop is smaller. Therefore, we may conclude that parallel flow
seems to be better.


\begin{center}
\includegraphics[scale=0.7]{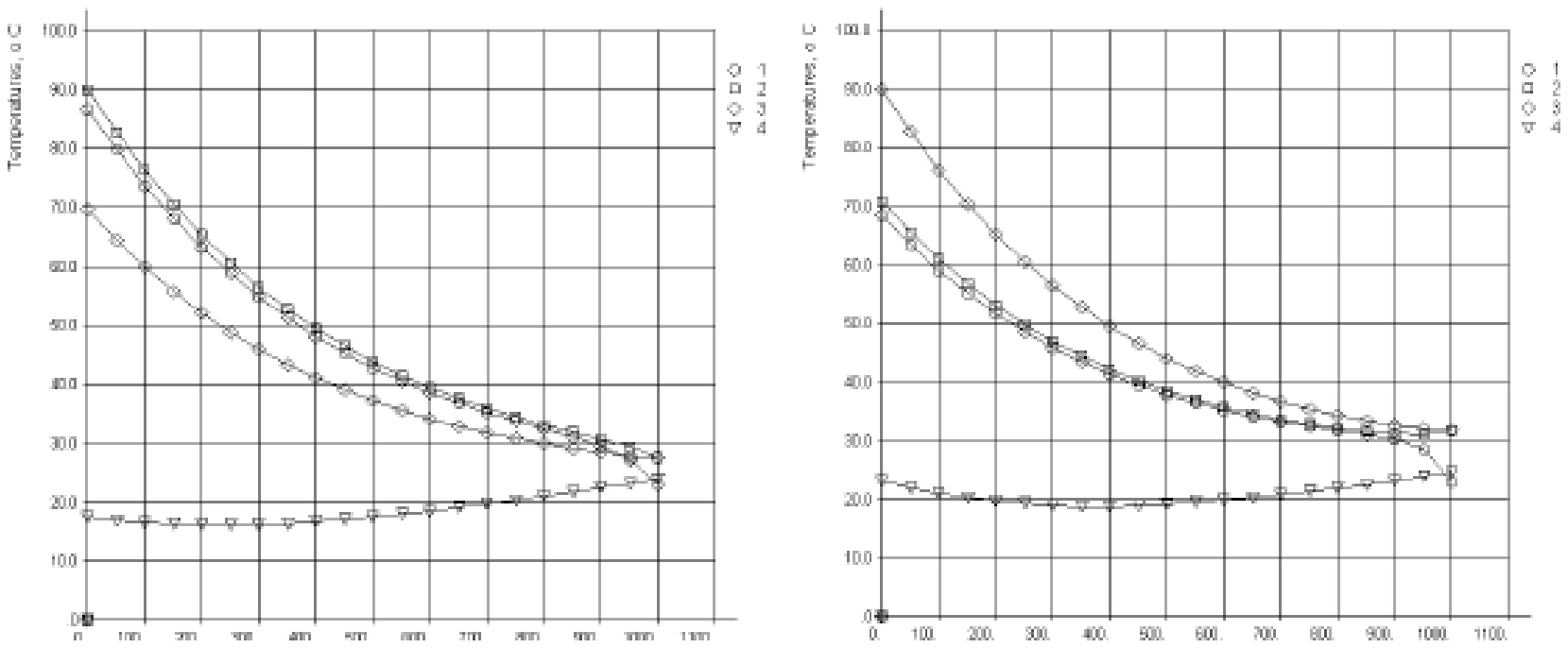}{\vskip 10pt} \small
{Figure 2. Temperature calculation, left for counter flow heat
exchange and right for parallel flow heat exchange.}\\ Legend:
1=$\square$ inner annulus, 2=$\Diamond$ outer annulus, 3=$\odot$
tubing, 4=$\triangledown$ earth
\end{center}

To conclude we point out that the linear model presented in this article 
has simplicity as its main advantage. It is not difficult to 
implement it in a computer code and it gives certain {\em initial} estimate 
of heat exchange in the system. Yet, important physical processes,
such as dissipation due to friction and variations of viscosities and
mass densities with the temperature, 
are not taken into account. They lead to nonlinear model that 
will be considered in our forthcoming publication.

\end{document}